# A New Greedy Randomized Adaptive Search Procedure For Multiobjective RNA Structural Alignment

Abdesslem Layeb, Amira Boudra, Wissem Korichi, Salim Chikhi

MISC Lab., Computer Science Department, University of Constantine, Ali Mendjli, Constantine 25000, Algeria
E-mail: layeb@umc.edu.dz, Boudra@umc.edu.dz,
Korichi@umc.edu.dz, Chikhi@umc.edu.dz

## ABSTRACT

*RNA secondary structures prediction is one of the main issues in bioinformatics. It seeks to elucidate structural conserved regions within a set of RNA sequences. Unfortunately, finding an accurate conserved structure is a very hard task to do. Within the present study, the prediction problem is considered as a multi objective optimization process in which the structural conservation and the sensitivity of the multiple alignment are optimized. The proposed method called GRASPMORSA is based on an aggregate function and GRASP procedure. The initial solutions are obtained by using a random progressive local/ global algorithm, and then they are refined by an iterative realignment. Experiments within a large scale of data have shown the efficacy and effectiveness of the proposed method and its capacity to reach good quality solutions.*

## 1. INTRODUCTION

Over the years, biologists have collected a great mass of information concerning nucleotide and protein sequences. Unfortunately, processing and interpretation of the content of these sequences have become very painful. For example a biologist must compare hundreds or even thousands of sequences to take their characteristics and the relationship between them. Unfortunately, a problem is posed: the task is even more difficult to accomplish if we take into account that the size of a single sequence can reach billions of bases. This is the case of the human genome. For this reason, the introduction of powerful tools and effective methods for the analysis and interpretation of biological data is highly requested. One of these very important problems found in various fields such as pharmacology, is the prediction of RNA secondary structures. The ribonucleic acids (RNA) are among the molecules stimulating the interest of the biologists. The RNA is now regarded as a potential target that is believed to be very interesting in pharmacology. In fact, RNA plays a multiple and a fundamental role in all cellular processes [1].The RNA plays a direct role in the catalytic processes like the synthesis of proteins. It plays also a fundamental role in regulation processes of the DNA replication, DNA transcription and translation [2]. There are close connections between RNA structures and their catalytic function. Indeed, it has been showed that a catalytic RNA becomes functional only when it has adopted its structure.

Consequently, it is very important to know the secondary structure and the possible tertiary structure of RNA molecules [3, 4].

Considering obtaining the structures of large RNA molecules by using nuclear magnetic resonance (NMR) spectrum is often difficult, the reliable forecast of RNA structures of their primary sequences is strongly required. Two main approaches are currently used to predict RNA secondary structures. The first is the comparative sequence analysis [5]. The basic idea is to examine homologous sequences to identify potential helices which maintain complementarities in sequences. The second approach is the thermodynamic optimization. In this method we use thermodynamics to determine structures with minimum or near minimum free energies [6].

The present paper is based on an idea of hybridization between thermodynamics and comparative approaches to address the RNA prediction problem. In order to obtain an efficient prediction of the RNA secondary structure with good alignment sensitivity, the optimization of both the Minimum Folding Energy (MFE) and the Weighted Sum of Pairs Score (WSPS) is recommended. Consequently, the problem becomes a multi objective optimization problem where conflicted objective functions are considered to be optimized [7]. Among the techniques used to solve the multi objective optimization problems: the aggregation method which combines the objective functions is considered to be new, each objective function is added in the new function with some weight [8]. The proposed hybrid approach is based on the metaheuristic GRASP (Greedy Randomized Adaptive Procedure). The first phase consists of constructing a multiple sequence alignment using a randomized greedy algorithm [9], while the second phase is to refine the solution found in the first by using an iterative realignment or simulated annealing. To validate our approach, we conducted tests on the base of benchmarks "BRALiBASE" which contain sets of RNA structures aligned tests manually created by biologists [10, 11]. For statistical validation of our results, we have used the Friedman test. The obtained results are very encouraging and prove the feasibility of the proposed method.

The rest of the paper is devoted to the mathematical formulation of the problem and a detailed presentation on the validation of the proposed approach. The paper ends with concluding statements.

## 2. MATHEMATICAL FORMULATION OF THE PROBLEM

The RNA structural alignment based on hybrid method requires two kinds of mathematical formulations: the formulation of the RNA secondary structure and the mathematical formulation of the multiple sequence alignment.

### 2.1. Mathematical formulation of RNA secondary structure

The secondary structure of RNA sequence is a set S of base pairs $(r_i, r_j)$ over the alphabet {A, C, G, U} satisfying the following criteria [12]:

1. $\forall (r_i, r_j) \in S (r_i, r_j) \in \{(A,U),(U,A),(G,C),(C,G),(G,U),(U,G)\}$
2. $1 \leq i \leq j \leq |S|$
3. $\forall (r_i, r_j), (r_i', r_j') \in S, i = i' \Leftrightarrow j = j'$
4. $(r_i, r_j) \in S \Rightarrow |j - i| \geq 4$

## 2.2. Mathematical formulation of multiple sequence alignment

Let $S = \{s_1, s_2, ..., s_n\}$ a set of *n* sequences with $n \geq 2$. Each sequence $s_i$ is a string defined over an alphabet A = {A, C, G, U}. The lengths of the sequences are not necessarily the same. The problem of MSA can be defined by specifying implicitly a pair $(\Omega, C)$ where $\Omega$ is the set of all feasible solutions that is potential alignments and *C* is a mapping $\Omega \to R$ called score of the alignment. Each potential alignment (Figure 1) is viewed as a set $S' = \{s'_1, s'_2, ..., s'_n\}$ satisfying the following criteria [9]:

- Each sequence $s'_i$ is an extension of $s_i$ and is defined over the alphabet $A' = A \cup \{-\}$. The symbol "−" is a dash denoting a gap. Gaps are added to $s_i$ in a way the deletion of gaps from $s'_i$ leaves $s_i$.

- For all i, j length($s'_i$)=length($s'_j$).

- A score of an alignment $S'$ denoted by $C(S')$ is defined as: $C(S') = \sum_i \sum_j sim(s'_i, s'_j)$

where $sim(s'_i, s'_j)$ denotes some similarity between each pair of sequences $s'_i$ and $s'_j$.

```
C A T G C G A G T A - G T A G
C A T G - - - G T A - G T A G
C C T G - G A G T A C G T A G
C A T G - - A G - - C G T A G
```

Figure 1. Example of multiple sequence alignment.

## 3. OBJECTIVE FUNCTION

The multi-objective problem of multiple RNA structural alignment is a very difficult optimization problem view the complexity of the biological data. In addition, the nonexistence of an efficient objective function to evaluate and identify optimal alignments regardless of the nature of the aligned sequences increases the degree of difficulty of the problem. Unfortunately, no method is entirely effective to do so [10, 13, 14, 15].

An idea that seems very interesting is to handle the problem of multiple structural alignment of RNA as a multi-objective optimization problem, where different scoring functions are optimized simultaneously. The multi-objective optimization appears as a natural setting for this study, it allows not only the optimization of several objectives functions simultaneously, but also captures the best features of each one; the strengths of one will fill the imperfections of the other objective. Generally, in multi-objective optimization, the optimization process will provide a set of solutions through the good compromise between the objective functions. This has the advantage of providing more choice to biologists at the decision making stage.

In order to evaluate our approaches, we have chosen two objectives simultaneously. The first objective function which is the most important is the minimum free energy of alignment MFE (Minimum Free Energy). It determines the score of the alignment from the consensus sequence (Figure 2) that idealizes a given region of the RNA in which each position represents the most frequently encountered base. The second objective function is the Weighted Sum of Peer Score (WSPS). This function is widely used in the literature to evaluate multiple sequence alignment.

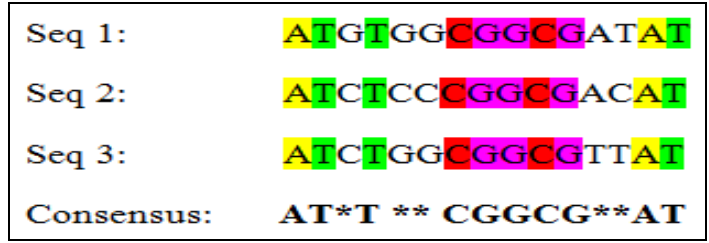

Figure 2. Example of consensus sequence

The objective behind the choice of these two functions is to evaluate the structural alignment according to the similarity and the structural conservation (WSPS and MFE), and therefore obtain an optimal solution by the method of aggregation. A multi-objective multiple RNA structural alignment can be formulated as follows:

$$\text{MOMRSA} = \begin{cases} \text{Max WSPS} \\ \text{Min MFE} \end{cases} \quad (2)$$

To combine these two objectives in a single aggregate function, we have transformed the minimization of the (1) formula into maximization by multiplying the objective function by -1. The MOMRSA problem can be reformulated as follows:

$$\text{MOMRSA} = \begin{cases} \text{Max WSPS} \\ \text{Max (-MFE)} \end{cases} \quad (3)$$

The aggregation function used is the following:

$$\text{MOMRSA} = \begin{cases} \text{Max WSPS} \\ \text{Max (-MFE)} \end{cases} \quad (4)$$

The aggregate function is given by the following formula:

$$\text{FMO} = \text{Max} \quad *\text{WSPS} - *\text{MFE} \quad (5)$$

and are weighted parameters, their sum is equal to one.

## 4. APPROACH GRASPMORSA : GRASP FOR THE MULTI OBJECTIVE   PREDICTION OF RNA STRUCTURES

The proposed approach, called GRASPMORSA, for solving the multi objective problem of multiple RNA structural alignment, is based on the GRASP procedure [16, 17]. The first phase consists to build an initial solution by using a new randomized heuristic. The second phase is used for refining the initial solution constructed in the first phase (Figure 3).

## 4.1. Construction of initial solutions

This phase builds gradually a multiple alignment of RNA sequences by using a randomized progressive approach. For this, we have adapted the new algorithm of progressive alignment proposed by Layeb et al., in [9] for the RNAs. This algorithm uses both the global alignment algorithm of Needleman-Wunsch and the local alignment algorithm of Smith-Waterman to build the progressive alignment. The local pairwise algorithm is used to align two sequences with large length difference in order to reduce the global misalignment, a threshold k is used to show that the difference between the sequences length is so large. However, in the case of two sequences with similar lengths, the global pairwise algorithm is more successful than the local algorithm. Our method uses an algorithm similar to the one proposed by Feng and Doolittle to build progressively a multiple sequence alignment [18]. We have used only sequence to sequence alignment and we have not used sequence to group or group to group alignments. The main novelty of our construction phase is its ability to produce a diverse set of good solutions; this behaviour is done by the insertion of some randomness in the progressive algorithm. Consequently, we could select with some probability a sequence even if it was not the closest one to the already aligned sequences. The global/local algorithm for the construction of an initial alignment is described as in algorithm 1 [19]:

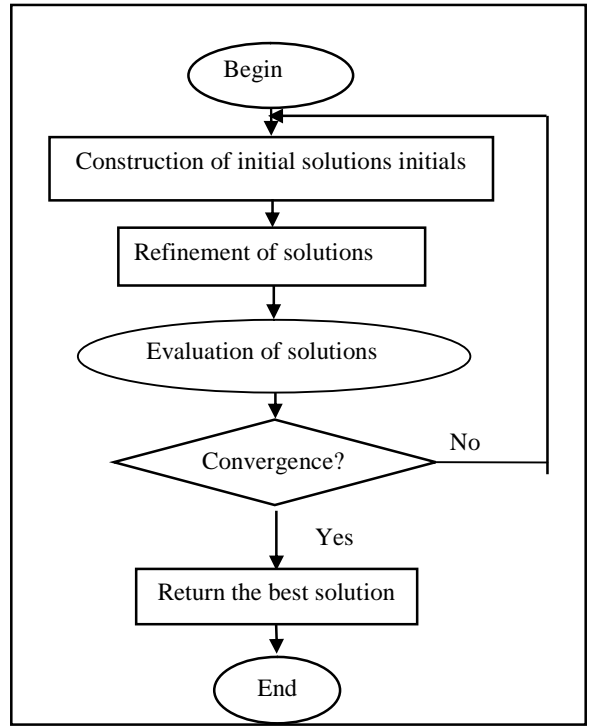

Figure 3. General structure of GRASP algorithm.

International Journal in Foundations of Computer Science & Technology (IJFCST), Vol. 3, No.1, January 2013

**Algorithm 1** : Global/local algorithm

**Input** : Set of sequences {$S_1$, $S_2$, …, $S_n$}
Begin
1. Construct the guide tree,
2. Choose the two closest sequences Si, Sj from the guide tree,
3. Compute the difference between the sequence lengths: $|length(S_i) - length(S_j)|$
4. <u>If</u>  $|length(S_i) - length(S_j)| > k$  <u>then</u> go to 5
    <u>Else</u> go to 6
5. Use a local pairwise algorithm to align the sequences $S_i$ et $S_j$
6. Use a global pairwise algorithm to align the sequences $S_i$ et $S_j$
7. Propagate gaps
8. <u>If</u>  (rand > 0,6) <u>then</u>
         Choose the next unaligned closest sequence $S_i$
      <u>otherwise</u>
         Choose a random unaligned sequence $S_i$
9. Choose the closest aligned sequence Si to Sj
10. <u>If</u> there is an unaligned sequence
       <u>Then</u>   go to 3
<u>End</u>
**Output :** Multiple sequence alignment

## 4.2 Refinement phase

Although the greedy approach of the first phase is quick and simple, as any other progressive approach, it may have many disadvantages. We can note for example that the quality of the progressive alignment depends on the choice of the first sequences to align. Thus, when the sequences are not very similar, the found alignment is less significant. Subsequently, the use of a refinement phase to improve the initial multiple alignment produced in the first phase is highly recommended. To improve the solutions of the first greedy phase, we have used a search local method based on iterative realignment (Figure 4). The new solution found in each step is evaluated by the aggregation function. The iteration continues until no more improvement can be done.

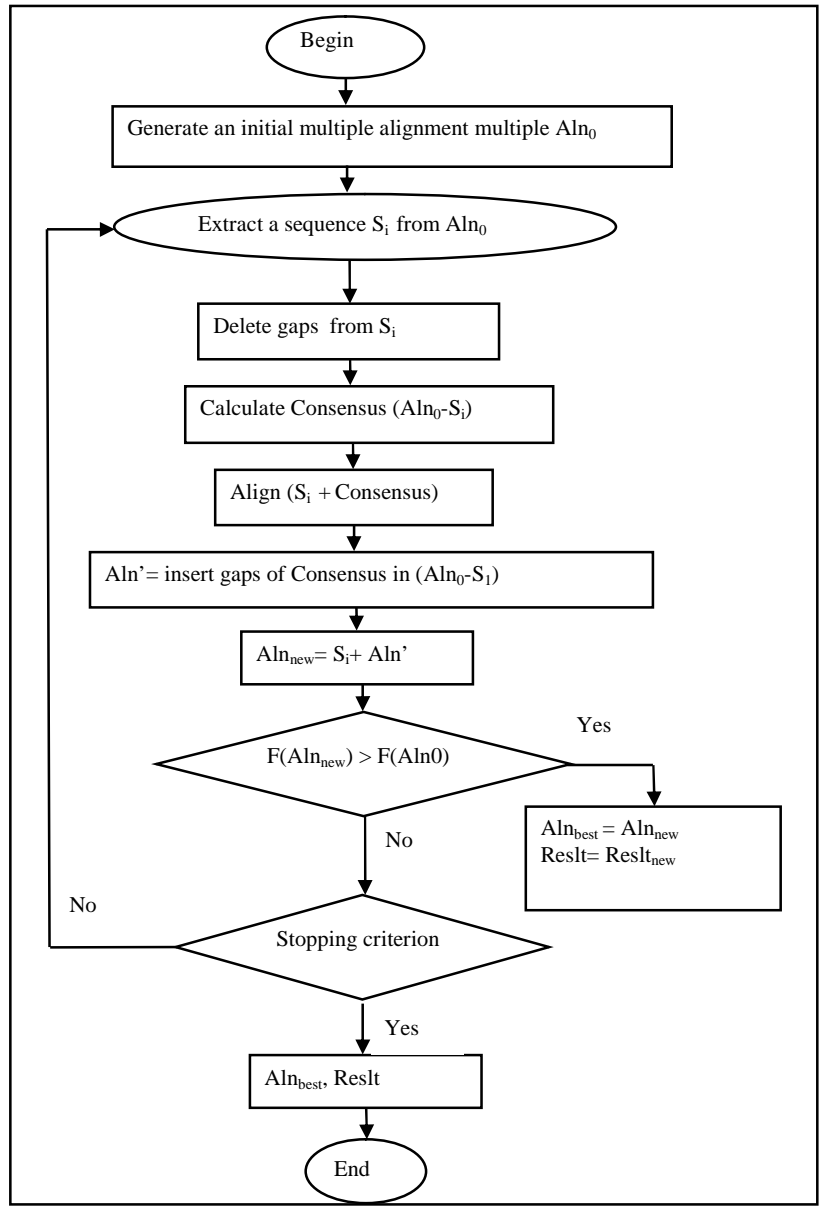

Figure 4. Structure of the solution refinement by iterative realignment.

## 5. IMPLEMENTATION AND EVALUATION OF THE PROPOSED APPROACH

The developed approach called GRASPMORSA has been implemented with MATLAB R2010a. Given that the most of the data of our approach are given in matrix form, we have chosen Matlab that has been used in many scientific computing domains (bioinformatics, optimization, signal processing, image processing, control, etc.), beside it has a bioinformatics toolbox containing various predefined bioinformatics functions. However, as an interpreted language, MATLAB is slower compared to other compiled languages such as C.

To evaluate the experimental performance of our approach, we have conducted several tests fromthe database benchmarks BRALiBASE II (http://projects.binf.ku.dk/pgardner/ bralibase/bralibase2.html) which contains sets of RNA structures tests, aligned manually by biologists. These tests are divided into five families of references: G2Introns, rRNA, SRP, tRNA and U5. They are differentiated by the number and the size of the sequences to be aligned, and the homology's degree between them.

To measure the biological quality of the found results, two independents measures, SPS and RNAz, have been used. The first is the famous SPS score calculated by the *Bali_score* program which measures the sensitivity of the RNA structural prediction. The other measure, (the most important) is the Structure Conservation Index SCI calculated by the program RNAz (**http://www.tbi.univie.ac.at/~wash/RNAz**). This index gives the measure of conserved secondary structure in the alignment. Unlike the SPS score, this measure is independent of reference alignment. An SCI is close to zero, if there is no common RNA structure between different sequences. If SCI is close to 1, it means a good indication on the secondary structure of the aligned sequences. However an SCI score > 1 indicates the existence of an RNA secondary structure that is supported by a compensatory preservation and / or shaped mutation of common structures. We note that the SCI marks only the accuracy of the alignment in terms of secondary structures information [10, 11]. Finally, the results are statistically validated using the Friedman tests.

## 6. RESULTS AND DISCUSSION

We have evaluated our approach (GRASPMORSA) by using the benchmark BraliBase. The results of our approach GRASPMORSA are described in tables (1 and 2). They clearly indicate the improvement of the aggregate function value, SPS and RNAz after refinement. It should be noted that in some cases, the orientation of both mathematical and biological measures is not always the same; this is caused by the difficulty of choosing the best parameters of the aggregate function and the MSA parameters. Figure 5 shows the Friedman test (0.05) for the SPS score that indicates clearly the amelioration after the refinement. Figure 6 shows the Friedman test (0.05) for MFE that shows a significant difference between the initial and the final MFE values after the refinement process. The rightmost position of the final value shows the required minimization.

Table 1. Result of the GRASPMORSA aggregation function.

| Tests | | WSPS | | MFE | | -0,7*MFE+0,3*WSPS | |
|---|---|---|---|---|---|---|---|
| | | Before | After | Before | After | Before | After |
| g2intron | aln1 | 4.1191 | 32.5642 | -35.7 | -35.8 | 26.2257 | 34.8293 |
| | aln11 | 13.1209 | 38.0263 | -48.7 | -48.7 | 38.0263 | 45.4979 |
| | aln13 | 7.2742 | 27.0191 | -28.2 | -34.5 | 21.9222 | 32.2557 |
| | aln20 | 9.3598 | 33.0479 | -43.2 | -43.2 | 33.0479 | 40.1544 |
| | aln27 | 15.1348 | 39.8746 | -40.9 | -48.0 | 33.1704 | 45.5624 |
| | aln51 | 3.3115 | 33.4824 | -38.9 | -46.1 | 28.2234 | 42.3147 |
| rRNA | aln10 | 20.4241 | 21.7389 | -60.1 | -61.2 | 48.1972 | 49.3617 |
| | aln12 | 16.1970 | 49.3001 | -61.4 | -63.8 | 47.8391 | 59.4500 |
| | aln16 | 21.3789 | 43.5837 | -53.1 | -53.1 | 43.5837 | 50.2451 |
| | aln30 | 24.1412 | 43.0824 | -51.2 | -51.2 | 43.0824 | 48.7647 |
| | aln34 | 21.4835 | 43.1295 | -49.1 | -52.4 | 40.8151 | 49.6188 |
| | aln50 | 40.5042 | 40.5042 | -39.9 | -39.9 | 40.0813 | 40.0813 |
| tRNA | aln25 | -1.2947 | 2.5798 | -22.6 | -27.5 | 15.4316 | 20.0239 |
| | aln34 | 27.7144 | 34.1198 | -35.8 | -36.9 | 33.3743 | 36.0659 |
| | aln50 | 22.7920 | 32.1076 | -36.1 | -36.1 | 32.1076 | 34.9023 |
| | aln60 | 23.8091 | 40.2722 | -40.8 | -47.2 | 35.7027 | 45.1216 |
| | aln81 | 31.0303 | 31.967 | -15.6 | -16.5 | 20.2291 | 21.1401 |
| | aln90 | 33.5619 | 35.388 | -15.2 | -16.5 | 20.7086 | 22.1664 |
| U5 | aln25 | 14.7383 | 33.3837 | -23.8 | -40.8 | 21.0815 | 38.5751 |
| | aln30 | 14.9081 | 31.3524 | -38.4 | -38.4 | 31.3524 | 36.2857 |
| | aln44 | 15.3976 | 34.0683 | -40.1 | -42.5 | 32.6893 | 39.9705 |
| | aln45 | 6.5713 | 30.4879 | -38.2 | -38.5 | 28.7114 | 36.0964 |
| | aln51 | 13.1871 | 35.5961 | -45.2 | -45.2 | 35.5961 | 42.3188 |
| | aln101 | 28.9837 | 46.2851 | -53.7 | -53.7 | 46.2851 | 51.4755 |

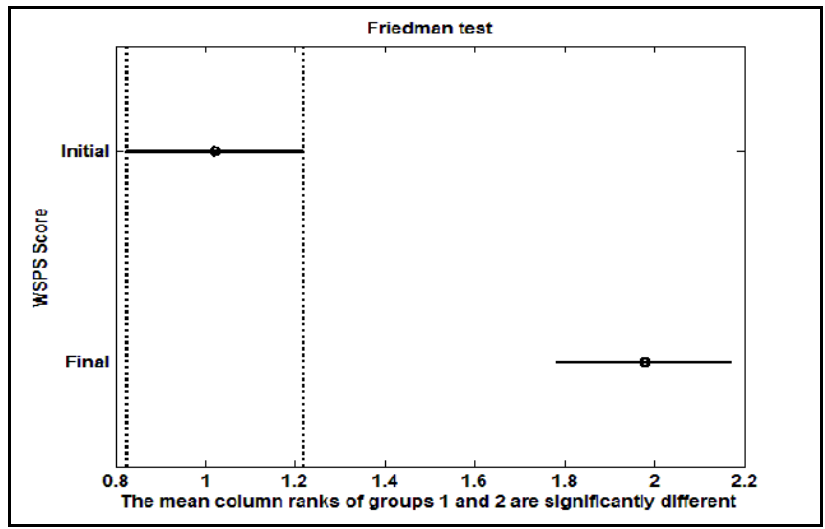

Figure 5. Friedman test (0.05) for WSPS

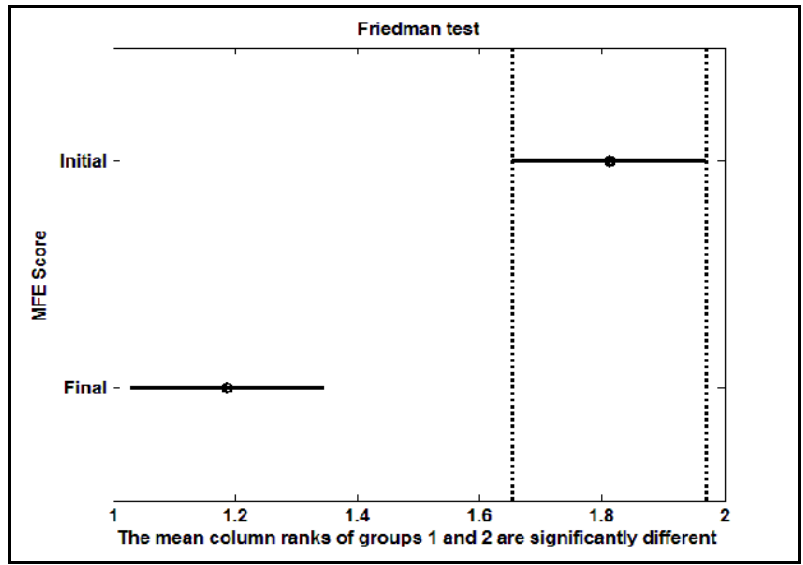

Figure 6. Friedman test (0.05) for MFE

Figure 7 presents the Friedman test (0.05) for the aggregate objective function. It clearly illustrates the improvement of values of aggregate function after refinement.

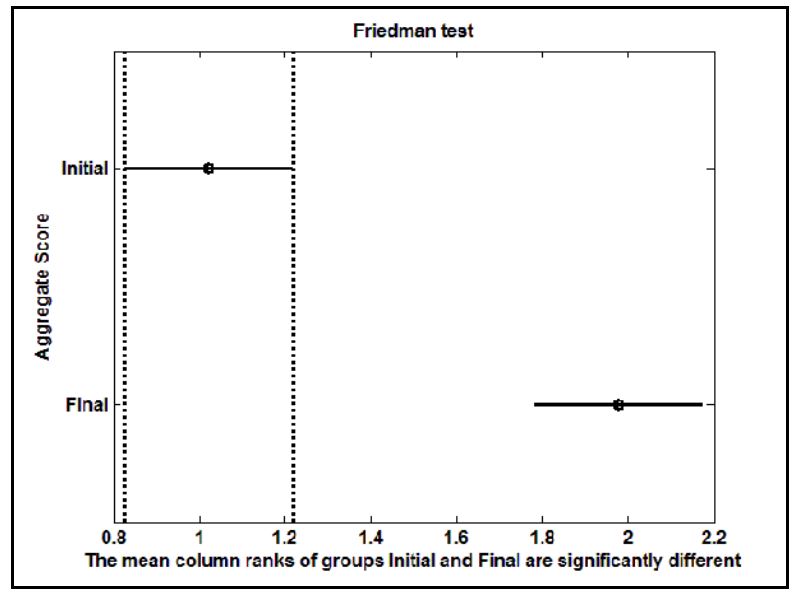

Figure 7. Friedman test (0.05) for objective function

Figure 8 of the Friedman test (0.05) shows a clear improvement of the SCI score after refinement process. On the other hand, Figure 9 presents the Friedman test (0.05) for the SPS score; it shows an acceptable difference between the initial and the final SPS scores after the refinement process.

Table 2. Biological result of GRASPMORSA.

| Tests | | SCI | | SPS | |
|---|---|---|---|---|---|
| | | Before | After | Before | After |
| g2intron | aln1 | 0.45 | 0.83 | 0.504 | 0.504 |
| | aln11 | 0.62 | 0.62 | 0.625 | 0.677 |
| | aln13 | 0.61 | 0.69 | 0.647 | 0.668 |
| | aln20 | 0.64 | 0.64 | 0.745 | 0.745 |
| | aln27 | 0.55 | 0.80 | 0.667 | 0.73 |
| | aln51 | 0.22 | 0.37 | 0.408 | 0.298 |
| rRNA | aln10 | 0.52 | 0.78 | 0.894 | 0.933 |
| | aln12 | 0.60 | 0.60 | 0.889 | 0.883 |
| | aln16 | 0.68 | 0.68 | 0.896 | 0.932 |
| | aln30 | 0.63 | 0.63 | 0.925 | 0.925 |
| | aln34 | 0.89 | 0.89 | 0.987 | 0.99 |
| | aln50 | 0.98 | 0.98 | 0.997 | 0.997 |
| tRNA | aln25 | 0.13 | 0.33 | 0.586 | 0.613 |
| | aln34 | 0.67 | 0.81 | 0.908 | 0.908 |
| | aln50 | 0.78 | 0.78 | 0.903 | 0.903 |
| | aln60 | 0.85 | 0.96 | 0.915 | 0.924 |
| | aln81 | 0.91 | 0.99 | 0.406 | 0.406 |
| | aln90 | 0.87 | 1.01 | 0.406 | 0.406 |
| U5 | aln25 | 0.04 | 0.20 | 0.453 | 0.137 |
| | aln30 | 0.30 | 0.30 | 0.045 | 0.069 |
| | aln44 | 0.30 | 0.50 | 0.277 | 0.378 |
| | aln45 | 0.08 | 0.12 | 0.495 | 0.495 |
| | aln51 | 0.05 | 0.05 | 0.301 | 0.287 |
| | aln101 | 0.75 | 0.75 | 0.89 | 0.89 |

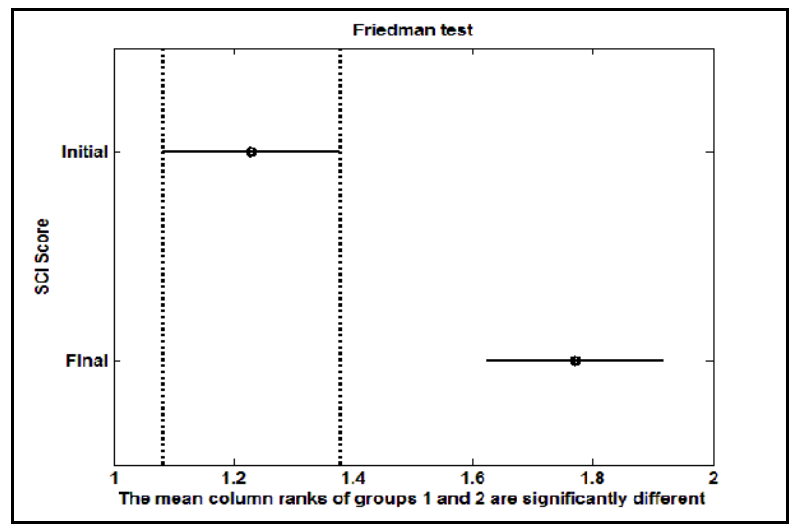

Figure 8. Friedman test (0.05) for SCI

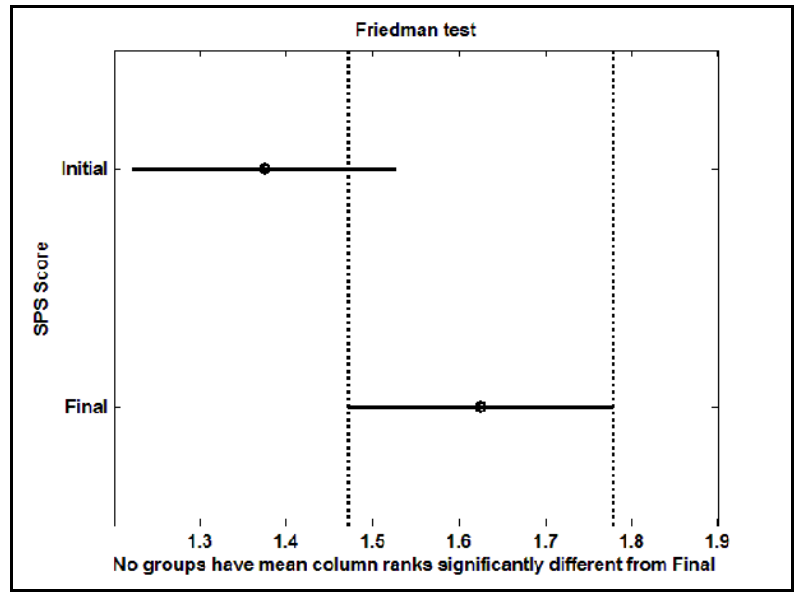

Figure 9. Friedman test (0.05) for SPS

In this section, we perform a comparative study between our program and the most used RNA programs in the literature. Table 3 summarizes the methods of multiple RNA structural alignment used in this experiment. The statistic studies relating to the RNA structural prediction of our approach are described in tables 4 and 5.

Table 3. Description of different comparison algorithms

| Nom | Description |
| --- | --- |
| **CLUSTAL** | Progressive algorithm for multiple sequence alignment |
| **DIALIGN** | Segment-based multiple sequence alignment |
| **MAFT** | Multiple Alignment using Fast Fourier Transform(FFT)/iterative algorithm |
| **PROALIGN** | A probabilistic multiple alignment program / progressive algorithme |
| **POA** | Multiple sequence alignment using partial order graphs |
| **HANDLE** | A probabilistic tool for creating and annotating multiple sequence alignments |
| **TCO_S** | Progressive algorithm for multiple sequence alignment |
| **MUSCLE** | Iterative algorithm for multiple sequence alignment |

As shown in the Friedman tests (figure 10 and 11), the obtained results are similar to most programs used in this experiment. For the SPS score (table 4, figure 10), our program is better than MAFT and has almost the same performances as TCO_S, POA, HANDLE and DIALIGN. But in terms of sensitivity, CLUSTAL, MUSCLE and PROALIGN are better than our program. For the SCI score (table 5, figure 11) which is more significant than that of SPS, GRASPMORSA and TCO_S are better than MAFFT, DIALIGN, POA and HANDLE but less efficient compared to CLUSTAL, PROALIGN and MUSCLE.

Table 4. Comparaison between GRASPMORSA and other methods (SPS)

| Tests | | SPS | | | | | | | | |
|---|---|---|---|---|---|---|---|---|---|---|
| | | GRASP MORSA | Clustal | DIA LIGN | MAFT | PRO ALIGN | POA | HANDLE | TCO_S | MUSCLE |
| g2intron | aln1 | 0.504 | 0.603 | 0.608 | 0.694 | 0.718 | 0.260 | 0.594 | 0.742 | 0.702 |
| | aln11 | 0.677 | 0.659 | 0.352 | 0.554 | 0.401 | 0.594 | 0.375 | 0.639 | 0.658 |
| | aln13 | 0.668 | 0.690 | 0.621 | 0.556 | 0.666 | 0.539 | 0.706 | 0.604 | 0.700 |
| | aln20 | 0.745 | 0.645 | 0.645 | 0.504 | 0.624 | 0.576 | 0.700 | 0.627 | 0.776 |
| | aln27 | 0.73 | 0.677 | 0.700 | 0.695 | 0.666 | 0.638 | 0.702 | 0.632 | 0.757 |
| | aln51 | 0.298 | 0.449 | 0.287 | 0.359 | 0.493 | 0.270 | 0.463 | 0.450 | 0.537 |
| rRNA | aln10 | 0.933 | 0.966 | 0.964 | 0.921 | 0.972 | 0.924 | 0.978 | 0.926 | 0.968 |
| | aln12 | 0.883 | 0.978 | 0.862 | 0.808 | 0.952 | 0.907 | 0.938 | 0.897 | 0.942 |
| | aln16 | 0.932 | 0.978 | 0.937 | 0.887 | 0.978 | 0.953 | 0.968 | 0.931 | 0.968 |
| | aln30 | 0.925 | 1.000 | 1.000 | 0.983 | 0.986 | 0.958 | 0.973 | 0.964 | 1.000 |
| | aln34 | 0.99 | 1.000 | 1.000 | 0.983 | 0.983 | 0.973 | 0.973 | 0.983 | 0.987 |
| | aln50 | 0.997 | 0.993 | 0.998 | 0.981 | 0.993 | 0.993 | 0.986 | 0.991 | 0.993 |
| tRNA | aln25 | 0.613 | 0.831 | 0.380 | 0.450 | 0.839 | 0.664 | 0.886 | 0.461 | 0.557 |
| | aln34 | 0.908 | 0.994 | 0.994 | 0.994 | 0.994 | 1.000 | 0.994 | 0.961 | 0.994 |
| | aln50 | 0.903 | 1.000 | 1.000 | 0.942 | 1.000 | 1.000 | 0.975 | 0.796 | 1.000 |
| | aln60 | 0.924 | 0.975 | 0.853 | 0.933 | 0.967 | 0.892 | 0.994 | 0.861 | 0.965 |
| | aln81 | 0.406 | 0.935 | 0.892 | 0.911 | 0.933 | 0.943 | 0.897 | 0.914 | 0.933 |
| | aln90 | 1.01 | 0.943 | 0.933 | 0.910 | 0.930 | 0.935 | 0.935 | 0.933 | 0.933 |
| U5 | aln25 | 0.137 | 0.751 | 0.674 | 0.648 | 0.736 | 0.658 | 0.705 | 0.707 | 0.773 |
| | aln30 | 0.069 | 0.639 | 0.612 | 0.660 | 0.696 | 0.599 | 0.731 | 0.647 | 0.743 |
| | aln44 | 0.378 | 0.839 | 0.743 | 0.575 | 0.860 | 0.849 | 0.791 | 0.622 | 0.835 |
| | aln45 | 0.495 | 0.588 | 0.627 | 0.553 | 0.671 | 0.557 | 0.539 | 0.616 | 0.729 |
| | aln51 | 0.287 | 0.558 | 0.634 | 0.541 | 0.733 | 0.519 | 0.705 | 0.532 | 0.729 |
| | aln101 | 0.89 | 0.876 | 0.893 | 0.871 | 0.913 | 0.904 | 0.913 | 0.878 | 0.904 |

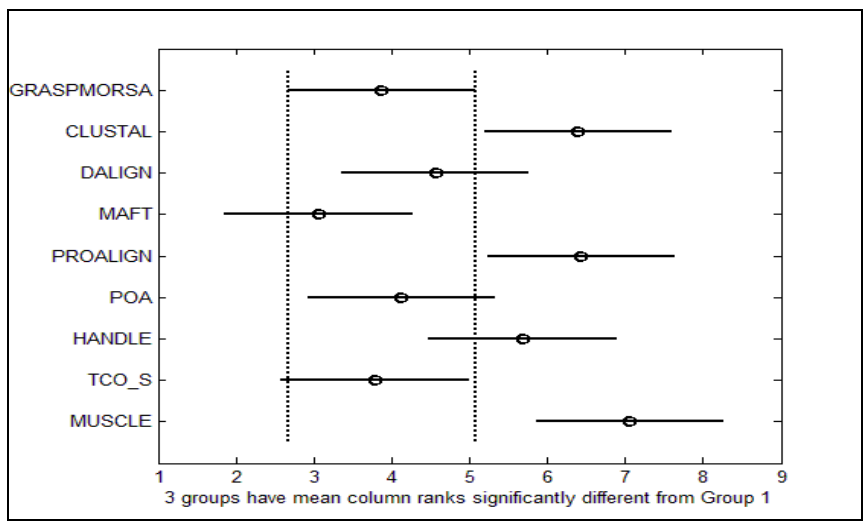

Figure 10. Friedman test (0.05) for SPS.

Table 5. Comparaison between GRASPMORSA and other methods (SCI)

| Tests | | SCI | | | | | | | | |
|---|---|---|---|---|---|---|---|---|---|---|
| | | GRASP MORSA | Clustal | DAI LIGN | MAFT | PRO ALIGN | POA | HANDLE | TCO_S | MUSCLE |
| g2intron | aln1 | 0.83 | 0.66 | 0.59 | 0.53 | 0.59 | 0.04 | 0.59 | 0.59 | 0.63 |
| | aln11 | 0.62 | 0.69 | 0.69 | 0.54 | 0.69 | 0.69 | 0.69 | 0.67 | 0.67 |
| | aln13 | 0.69 | 0.72 | 0.66 | 0.62 | 0.69 | 0.54 | 0.66 | 0.66 | 0.70 |
| | aln20 | 0.64 | 0.66 | 0.66 | 0.00 | 0.58 | 0.64 | 0.60 | 0.63 | 0.66 |
| | aln27 | 0.80 | 0.58 | 0.58 | 0.59 | 0.58 | 0.58 | 0.56 | 0.58 | 0.57 |
| | aln51 | 0.37 | 0.38 | 0.26 | 0.19 | 0.31 | 0.23 | 0.33 | 0.54 | 0.47 |
| rRNA | aln10 | 0.78 | 0.88 | 0.79 | 0.75 | 0.91 | 0.80 | 0.86 | 0.92 | 0.98 |
| | aln12 | 0.60 | 0.70 | 0.37 | 0.22 | 0.68 | 0.58 | 0.64 | 0.60 | 0.60 |
| | aln16 | 0.68 | 1.03 | 0.82 | 0.56 | 1.03 | 0.93 | 0.87 | 0.82 | 0.80 |
| | aln30 | 0.63 | 1.09 | 1.09 | 1.02 | 1.01 | 0.98 | 0.92 | 1.02 | 1.09 |
| | aln34 | 0.89 | 0.92 | 0.92 | 0.90 | 0.90 | 0.88 | 0.87 | 0.90 | 0.90 |
| | aln50 | 0.98 | 0.97 | 0.95 | 0.98 | 0.98 | 0.98 | 0.95 | 0.96 | 0.98 |
| tRNA | aln25 | 0.33 | 0.94 | 0.07 | 0.20 | 1.06 | 0.24 | 1.07 | 0.11 | 0.51 |
| | aln34 | 0.81 | 1.09 | 1.09 | 1.09 | 1.09 | 1.09 | 1.09 | 1.09 | 1.09 |
| | aln50 | 0.78 | 1.08 | 1.08 | 0.64 | 1.08 | 1.08 | 1.05 | 0.43 | 1.08 |
| | aln60 | 0.96 | 1.16 | 0.79 | 0.96 | 1.16 | 1.00 | 1.16 | 0.93 | 1.16 |
| | aln81 | 0.99 | 1.05 | 0.83 | 0.94 | 1.09 | 1.05 | 0.81 | 1.05 | 1.05 |
| | aln90 | 1.01 | 1.00 | 1.01 | 0.93 | 1.03 | 1.01 | 1.00 | 1.01 | 1.01 |
| U5 | aln25 | 0.20 | 0.28 | 0.23 | 0.19 | 0.28 | 0.12 | 0.11 | 0.26 | 0.32 |
| | aln30 | 0.30 | 0.26 | 0.25 | 0.26 | 0.52 | 0.23 | 0.50 | 0.27 | 0.49 |
| | aln44 | 0.50 | 0.50 | 0.06 | 0.13 | 0.50 | 0.50 | 0.50 | 0.50 | 0.51 |
| | aln45 | 0.12 | 0.10 | 0.05 | 0.00 | 0.14 | 0.05 | 0.05 | 0.13 | 0.05 |
| | aln51 | 0.05 | 0.16 | 0.01 | 0.12 | 0.17 | 0.10 | 0.00 | 0.04 | 0.11 |
| | aln101 | 0.75 | 0.88 | 0.74 | 0.74 | 0.81 | 0.81 | 0.81 | 0.80 | 0.77 |

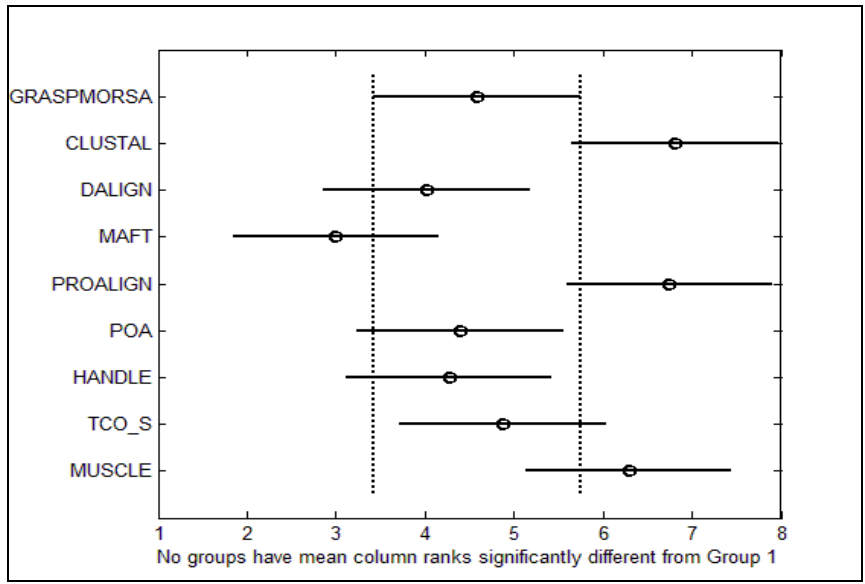

Figure 11. Friedman test (0.05) for SCI.

## 7. CONCLUSION

To solve the multi-objective problem of the RNA secondary structures prediction, we have presented in this paper an approach called GRASPMORSA based on the metaheuristic GRASP and the aggregation multiobjective optimization technique. The proposed metaheuristic is basically composed of two phases. The initial phase is a greedy randomized approach for constructing the initial solutions based on new randomized progressive multiple sequence alignment. The second phase is a refinement procedure used for improving the solutions produced in the first phase. To perform this refinement, a local search method based on the realignment technique is used. In order to evaluate each new solution, an aggregate function based on both SPS and MFE score functions is used. The *BraliBase* benchmark base is used to evaluate the performance of the proposed approach. The found results are very encouraging; they are comparable to the most effective methods of multiple RNA structure prediction. Finally, there are several ways to improve the proposed approach like the use of more sophisticated local search methods like tabu search or simulated annealing. On the other hand, the use of the Pareto-based multiobjective optimization methodology can increase considerably the performance of the proposed algorithm.

## AUTHORS

**Abdesslem Layeb** is an Associate Professor in the Department of Computer Science at the University of Constantine, Algeria. He is a member of MISC Laboratory. He received his PhD in Computer Science from the University Mentouri of Constantine, Algeria. He is interested by combinatorial optimisation methods and their applications to solve several problems from different domains like bioinformatics, imagery, formal methods, etc.

**Amira Boudra** received Master's degree from the University of Constantine, Algeria in 2011. Her main interests are the bioinformatics and the combinatorial optimization methods.

**Wissem Korichi** received Master's degree from the University of Constantine, Algeria in 2011. Her main interests are the bioinformatics and the combinatorial optimization methods.

**Salim Chikhi** received his PhD in Computer Science from the University of Constantine, in 2005. He is currently a Professor at the Mentouri University, Constantine, Algeria. He is the Leader of the SCAL team of MISC Laboratory. His research areas include soft computing and artificial life techniques and their application in several domains.